\documentclass[twocolumn,twoside,slac_two]{revtex4}
\usepackage{graphicx}
\usepackage{fancyhdr}
\newcommand{\fer}{{\it Fermi}}
\newcommand{\wse}{{\it WISE}}
\newcommand{\bzcat}{ROMA-BZCAT}
\begin{document}

\title{A new method for the extraction of mid-infrared $\gamma$-ray emitting candidate blazars}

%

\author{R. D'Abrusco}
\affiliation{Harvard-Smithsonian Center for Astrophysics, 60 Garden 
		Street, Cambridge, MA, 02138, US}

\author{F. Massaro}
\affiliation{SLAC National Laboratory and Kavli Institute for Particle Astrophysics and Cosmology, 2575 
Sand Hill Road, Menlo Park, CA 94025, USA}

\author{A. Paggi}
\affiliation{Harvard-Smithsonian Center for Astrophysics, 60 Garden 
		Street, Cambridge, MA, 02138, US}
		
\author{N. Masetti}		
\affiliation{INAF/IASF di Bologna, via Gobetti 101, I-4019 Bologna, Italy}

\author{G. Tosti}
\affiliation{Dipartimento di Fisica, Universit\`a degli Studi di Perugia, 06123 Perugia, Italy}

\author{M. Giroletti}
\affiliation{INAF Istituto di Radioastronomia, via Gobetti 101, 40129, Bologna, Italy}

\author{H.~A. Smith}
\affiliation{Harvard-Smithsonian Center for Astrophysics, 60 Garden 
		Street, Cambridge, MA, 02138, US}

\begin{abstract}

We present a new method for identifying blazar candidates by examining the \emph{locus}, 
i.e. the region occupied by the \fer\ $\gamma$-ray blazars in the three-dimensional 
color space defined by the WISE infrared colors.
This method is a refinement of our previous approach that made use of the 
two-dimensional projection of the distribution of WISE $\gamma$-ray emitting blazars 
(the \emph{Strip}) in the three WISE color-color planes~\citep{massaro12a}.
In this paper, we define the three-dimensional \emph{locus} by means of a 
Principal Component (PCs) analysis of the colors distribution of a
large sample of blazars composed by all the \bzcat\ sources with counterparts in the WISE 
All-Sky Catalog and associated to $\gamma$-ray source in the second \fer\ LAT catalog (2FGL) 
(the WISE \fer\ Blazars sample, WFB). Our new procedure, as reported in~\citep{dabrusco2013}, 
yields a total completeness of $c_{\mathrm{tot}}\!\sim$81\% and total efficiency of 
$e_{\mathrm{tot}}\!\sim$97\%.

\end{abstract}

\maketitle

\thispagestyle{fancy}


\section{Introduction}

Unveiling the nature of the Unidentified Gamma-ray Sources (UGS) 
is one of the main scientific objectives of the ongoing \fer\ $\gamma$-ray mission.
Recently, several attempts have been performed to associate or characterize the UGSs, either 
using X-ray observations \citep[e.g.,][]{mirabal09a,mirabal09b} or with statistical 
approaches~\citep[e.g.][]{mirabal10,ackermann12}. Nevertheless, according to~\cite{nolan12}, 
31\% of the $\gamma$-ray sources in the second \fer\ LAT catalog (2FGL) remain unidentified
and many of these unidentified sources could be blazars,
since blazars are known to dominate the $\gamma$-ray sky~\citep[e.g.][]{hartman99,abdo10a},
and among the 1297 associated sources within the 2FGL, 805 (62\%) are known blazars~\cite{nolan12}.
Therefore it is important to devise an efficient means of identifying candidate blazars 
among these sources.

Blazars come in two main classes: the BL Lac objects, 
which have featureless optical spectra, and the more luminous 
Flat-Spectrum Radio Quasars which, typically, show prominent optical spectral emission
lines \citep{stickel91}.
In the following discussion, we label the BL Lac objects as BZBs and the 
Flat-Spectrum Radio Quasars as BZQs, following the nomenclature 
of the Multi-wavelength Catalog of blazars \citep[\bzcat,][]{massaro09}.

Using the preliminary data release of the Wide-field Infrared Survey Explorer (\wse) 
\citep[see~][for more details]{wright10}\footnote{http://wise2.ipac.caltech.edu/docs/release/prelim/}, 
we showed that the $\gamma$-ray blazar population occupies a distinctive region 
of the WISE color parameter space (called the WISE Gamma-ray \emph{Strip}
\cite{massaro11a,dabrusco12}). Taking advantage of the 
much larger data set now available thanks to the WISE 
All-Sky archive\footnote{http://wise2.ipac.caltech.edu/docs/release/allsky/}, 
released in March 2012, in this work we present a revisited definition of 
the region occupied by the $\gamma$-ray blazars. 
We will refer to the 3-dimensional region occupied by the $\gamma$-ray emitting blazars 
as the \emph{locus} while we will continue to indicate the 2-dimensional projection of the 
\emph{locus} in the $[3.4]\!-\![4.6]$ vs $[4.6]\!-\![12]$ $\mu$m color-color plane  as WISE Gamma-ray 
\emph{Strip}.
In this proceeding we determine a geometrical description of the WFB \emph{locus} in the space generated 
by the Principal Components of their distribution in the WISE color space and apply our association 
procedure the whole sample of blazars that belong to the 2FGL.

\section{The WISE {\it Fermi} Blazars sample}

We found 3032 out of 3149 (i.e., 96.3\% of the \bzcat) blazars with an IR counterpart 
within 3.3$''$ in the WISE All-Sky data archive.  
In this sample, there are only 2 multiple matches out of 3032 spatial associations, 
for which we used the IR data of the closest WISE source in the following analysis. 
The probability of a chance associations for these 3032 is $\sim\!$ 3.3\%,
implying that $\sim\!$ 100 sources associated within the above radius could 
be spurious associations.

Of these 3032 blazars, 1172 are BZBs, including 919 BL Lacs and 253 BL Lac candidates,
1642 are BZQs and 218 are BZUs. It is also worth noticing that all the blazars 
associated between the \bzcat\ and the WISE all-sky data release 
are detected in the first two filters at 3.4 and 4.6 $\mu$m. Among the 3032 selected 
blazars, only 673 have a counterpart in the $\gamma$-rays according to  
the 2FGL and to the CLEAN sample presented in the second \fer\ LAT Catalog of active 
galactic nuclei \citep[2LAC;][]{ackermann11}. 637/673 (i.e., 94.7\%) of these blazars (333 BZBs, 277 
BZQs, and 27 BZUs) are detected in all four WISE bands. As in~\citep{massaro12b} the sample of $\gamma$-ray
emitting blazars in the \bzcat\ catalog was derived 
excluding the BZUs sources from our sample of $\gamma$-ray loud blazars. For this reason, the final
WFB sample includes only 610 WISE sources out of 673 WISE counterparts. We have used the WFB
sample to characterize the model of the \emph{locus} in the WISE color space.

\section{The locus parametrization}

\begin{figure}[] 
	\includegraphics[height=9cm,width=9cm,angle=0]{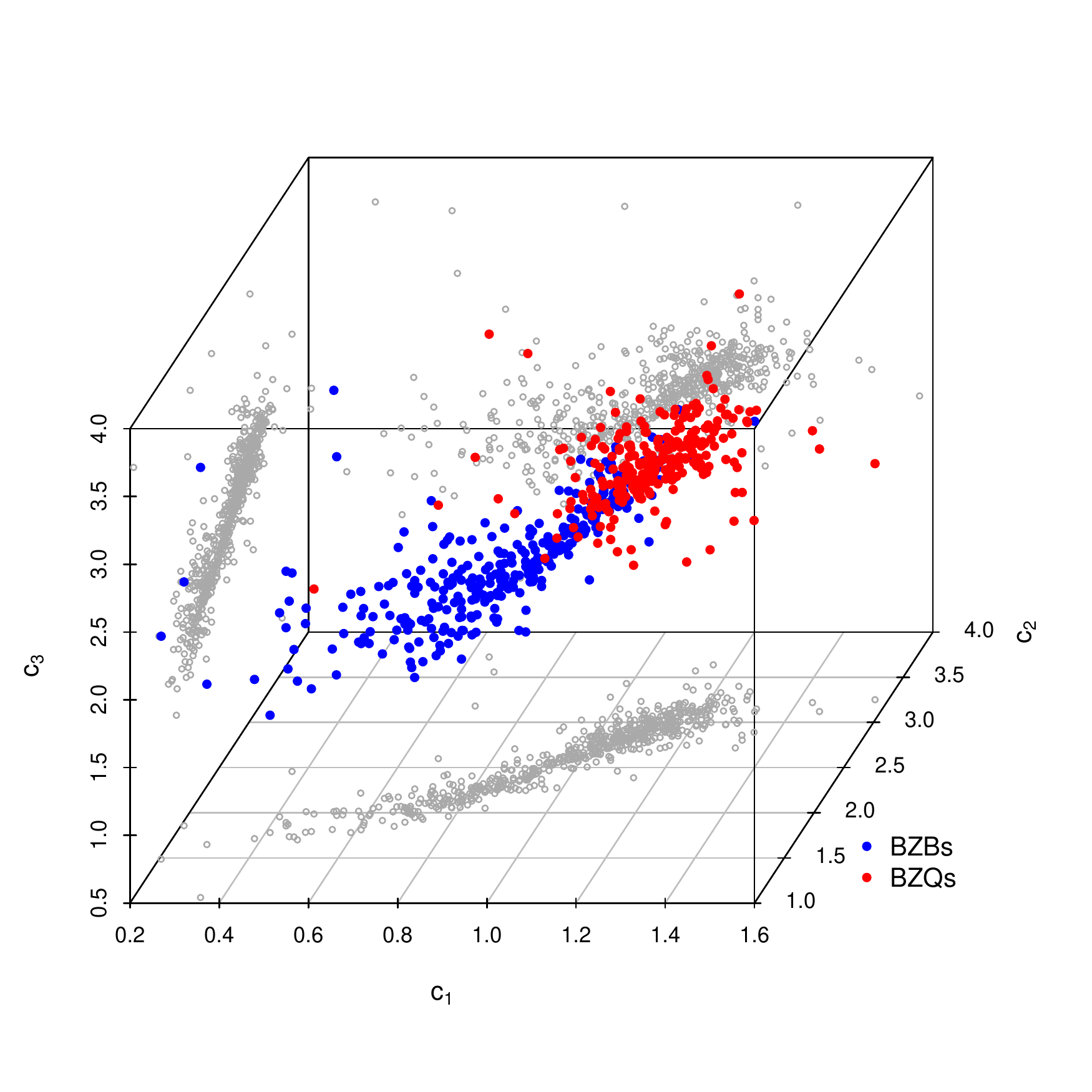}
	\caption{Scatterplot of the WFB sources in the three-dimensional WISE color space. The spectral 
	class of the WFB sources is color-coded, while the gray points represent the 
	projections of the WFB sample in the three color planes generated by WISE colors.}
          \label{fig:3d}
\end{figure}

The new parametrization of the WFB \emph{locus} is based on a new model of the \emph{locus} in the 
PCs space generated by the WISE colors of the WFB WISE counterparts, and 
on revised definition of the statistical quantity used to evaluate the compatibility of a generic WISE 
source with the \emph{locus} model, the score. The distribution of the WFB 
sources in the three-dimensional WISE color space is axisymmetric along a slew line 
(see Figure~\ref{fig:3d}), so that a simple geometrical description of the \emph{locus} can be 
determined in the PCs space. 

Principal Component Analysis (PCA) uses an 
orthogonal transformation to convert a set of observations of possibly correlated 
variables into a set of values of linearly uncorrelated variables, the PCs. This 
transformation $T\!:\!(c_{1},c_{2},c_{3})\rightarrow(\mathrm{PC}_{1},\mathrm{PC}_{2},
\mathrm{PC}_{3})$ is defined so that the first PC (PC$_{1}$) accounts for as much of the variance in 
the data as possible, and each following component (PC$_{2}$, PC$_{3}$, etc. up to 
the dimensionality of the initial space) has the highest variance possible under the constraint that it 
is orthogonal to the preceding components. 
In our case, the WFB sources in the three-dimensional PCs space based on their color distributions lie almost 
perfectly along the PC$_{1}$ axis and are distributed symmetrically in the PC$_{2}$ vs PC$_{3}$ 
plane around the PC$_{1}$ line. Based on the shape of the \emph{locus} in the PCs space, 
we choose to define its geometrical model using a cylindrical parametrization, 
with axis aligned along the PC$_{1}$ axis (see Fig~\ref{fig:scheme}). 
The \emph{locus}, as a whole, is modeled by three distinct cylinders: the first two of these cylindrical 
regions are dominated by BZB and BZQ sources respectively, while the third cylinder is defined as the 
region where the WFB population is \emph{mixed} in terms of spectral classes (the Mixed region, 
hereinafter). 

\begin{figure}[] 
	\includegraphics[width=65mm]{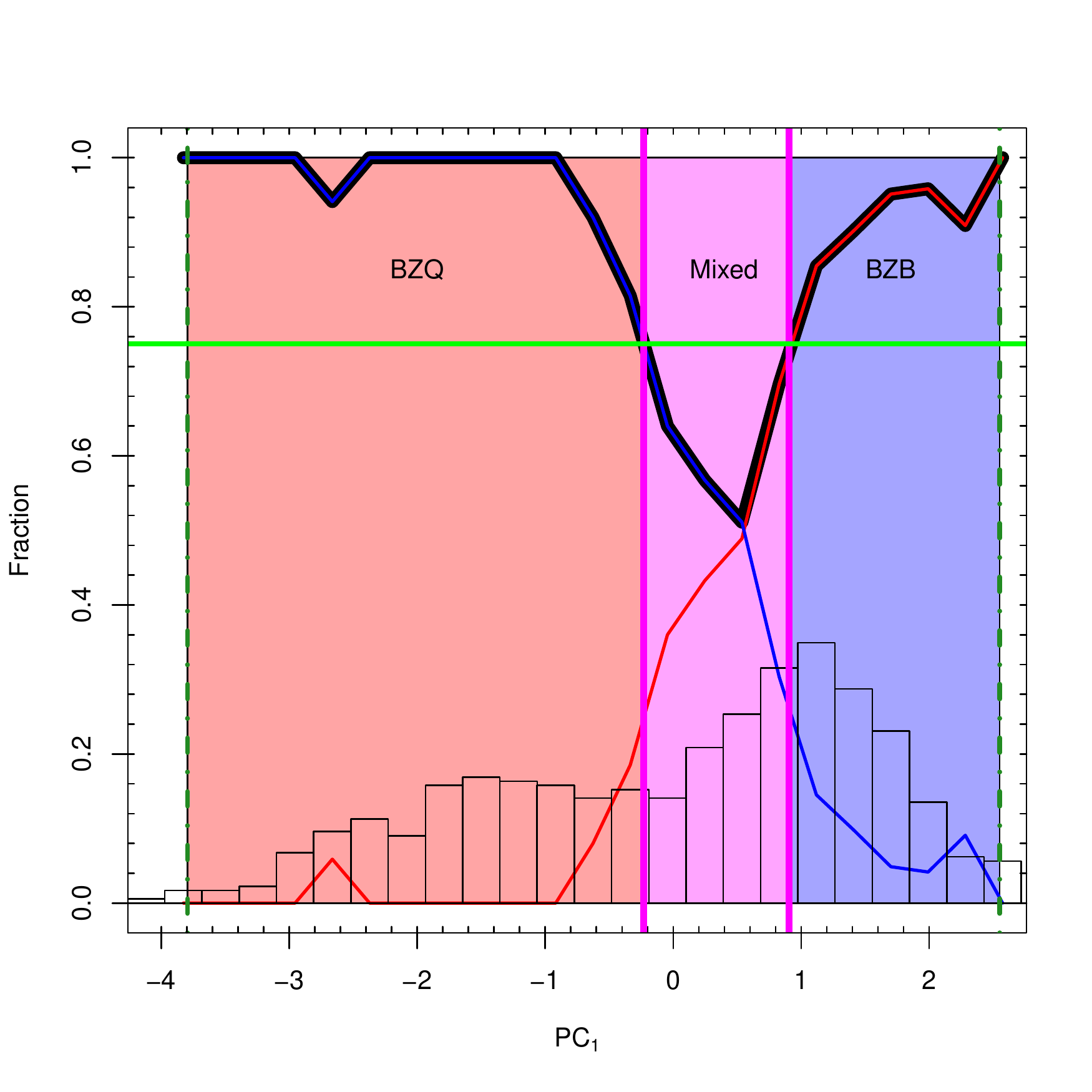}
          \caption{Boundaries of the three sections of the WFB \emph{locus} in the PCs space
          along the PC$_{1}$ axis. The solid black line represent the ``purity'' of the WFB population, 
          i.e. the fraction of the dominant spectral class relative to the other spectral class. The solid 
          red and blue lines represent the fraction of BZQs and BZBs sources, while the histogram 
          in the background represents the normalized density of the distribution of the whole WFB sample
          along the PC$_{1}$ axis. The horizontal green line shows the threshold used to determine 
          the boundaries of the mixed region.}
          \label{fig:cylinders}
\end{figure}

The upper and lower boundaries of the model along the 
PC$_{1}$ axis have been determined requiring that 90\% of the total number of WFB sources is contained 
within the boundaries of the cylinder, with 5\% of the sources outside of the boundaries of the model 
on each side of the model along the PC$_{1}$ axis. 
The boundaries of the Mixed section along the PC$_{1}$ axis have been defined by requiring 
that, in this region, the fraction of either spectral class is smaller than 80\% of the total number of WFB 
sources. The three boundaries along the PC$_{1}$ axis defining the three sections 
of the WFB \emph{locus} model are shown in Figure.~\ref{fig:cylinders}.

\noindent The variances of the distribution of the WFB distribution in the
PC space along the second and third PCs are $\sigma^2_{\mathrm{PC_{2}}}\!=\!0.61$ and 
$\sigma^2_{\mathrm{PC_{3}}}\!=\!0.58$ respectively. Based on this fact, we have 
modeled the bases of the cylinders as circles centered on the axis of the first principal component
PC$_{1}$ (the variance of the WFB distribution along PC$_{1}$ is 
$\sigma^1_{\mathrm{PC_{3}}}\!=\!1.53$).
The radii of the circular bases of each of the three cylinders representing the three
different sections of the WFB \emph{locus} in the PCs space have been determined independently as 
the radii containing the 90\% of the WFB sources in each section. The radii of each of the three 
cylinders are defined 
in the plane generated by the PC$_{2}$ and PC$_{3}$ axes and evaluate d as  
$R=\sqrt{\mathrm{PC}_{2}^2\!+\!\mathrm{PC}_{3}^2}$. 

\subsection{The score}
\label{subsec:score}

\begin{figure}[] 
	\includegraphics[width=65mm]{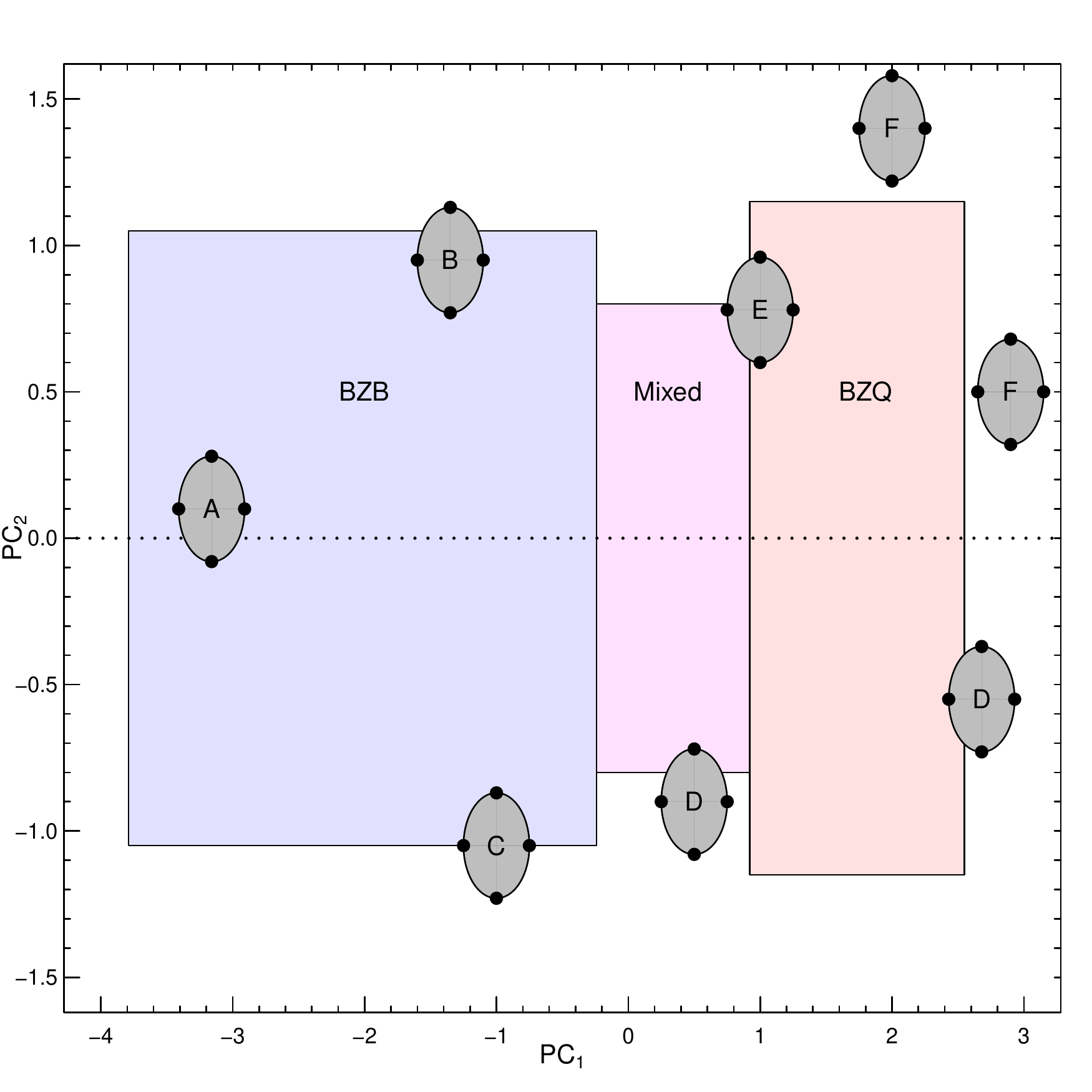}
           \caption{Schematic representation of the possible positions of the uncertainty ellipsoid
           of a generic WISE source in the PCs space relative to the cylindrical models of the WFB
           \emph{locus} (see descriptions of the different cases in the text).}
           \label{fig:scheme}
\end{figure}

The distance of a generic WISE source to the model of the WFB \emph{locus} in 
the PCs space can be evaluated quantitatively using a numeric quantity that we call the score. 
The generic WISE source with colors $(\tilde{c}_{1}, \tilde{c}_{2}, \tilde{c}_{3})$ can be projected onto
the PCs space by applying the orthogonal transformation determined by the PCA performed on 
the WFB sample for the modelization of the WFB \emph{locus} in the PCs space. 
Thus, the position of the generic WISE source in the PCs space is determined by the 
PCs values $(\tilde{\mathrm{PC}}_{1},\!\tilde{\mathrm{PC}}_{2},
\!\tilde{\mathrm{PC}}_{3})\!=\!T(\tilde{c}_{1},
\!\tilde{c}_{2},\!\tilde{c}_{3})$. To take into account the uncertainties on the values of the WISE colors,
the standard deviations on each color are also projected onto the PCs space and are used to 
define the error bars on the position of the source in the PCs space: $(\pm\sigma_{\tilde{\mathrm{PC}}_{1}},
\!\pm\sigma_{\tilde{\mathrm{PC}}_{2}},\!\pm\sigma_{\tilde{\mathrm{PC}}_{3}})\!=\!T(\pm\sigma_{\tilde{c}_{1}},
\!\pm\sigma_{\tilde{c}_{2}},\!\pm\sigma_{\tilde{c}_{3}})$. We simply assume that the 
generic WISE source is represented in the PCs space by the ellipsoid generated by the 
segments with extremes $\mathrm{PC}_{i} \pm \sigma_{\mathrm{PC}_{i}}$, hereinafter the 
uncertainty ellipsoid. Each of the six points at the extremes of the axes of the uncertainty 
ellipsoid in the PCs space will be generically called extremal point. The possible
positions of the uncertainty ellipsoid associated with a generic WISE source relative to each 
of the three cylinders of the \emph{locus} model
(schematically shown in Figure~\ref{fig:scheme} for one two-dimensional section of the PCs space) 
fall in one of the following cases: six extremal points within a cylinder (point A in Figure~\ref{fig:scheme});
five extremal points within a cylinder (point B in Figure~\ref{fig:scheme}); three extremal points 
within a cylinder (point C in Figure~\ref{fig:scheme}); one extremal point within a cylinder 
(points D in Figure~\ref{fig:scheme}); no extremal points within any cylinder (points F in 
Figure~\ref{fig:scheme}). Other combinations are not possible because the axes of the uncertainty 
ellipsoids are either parallel or orthogonal to the PC$_{1}$ axis of the PCs space. Points with
any number of extremal points within two cylinders (like point E in Figure~\ref{fig:scheme}) are assigned
a distinct score value for either cylinder according to the number of extremal points contained
in each one.
 			
\noindent The score $s$ for a generic WISE source with $n$ extremal points contained in one of the three
sections of the WFB \emph{locus} model is defined as:

\begin{equation}
	\nonumber
	\mathrm{s} = \frac{1}{6^{\phi}}\cdot\!n^{\phi}
	\label{eq:score}
\end{equation}

\noindent where $\phi$ is the \emph{index} of the score assignment law. 
This is a simple generalization of the most natural choice that would assign to
each extremal point within the \emph{locus} model 1/6, defining the total score of a
source as linearly proportional to the number of extremal points within the model cylinders. 
This behavior is obtained in the general equation when $\phi\!=\!1$. 
Changing the value of $\phi$ is useful to tweak the performances of the association 
procedure in terms of the purity and completeness of the final sample of candidate blazars.

So far, the score assigned to a generic WISE source can take one of six different 
values determined by the score assignment law in Equation~\ref{eq:score}. To 
penalize the WISE sources with large uncertainties on the observed colors (and, 
in turn, large volume of the uncertainty ellipsoid in the PCs space) relatively to 
other WISE sources with the same number of extremal points contained in the 
\emph{locus} model but smaller errors, we multiply the score obtained using 
Eq.~\ref{eq:score} by the
ratio of the absolute values of the logarithms of the volume of the uncertainty 
ellipsoid of the source considered and of the volume of the largest uncertainty ellipsoid
for WFB sources. Thus, for each of the three regions of the \emph{locus} model, the 
weighted score is defined as:

\begin{equation}
	\nonumber
	\mathrm{s}_{w} = \mathrm{s}\!\cdot\frac{\|\log{V}\|}{\|\log{(\mathrm{max(V_{\mathrm{WFB}})})}\|}
	\label{eq:weiscore}
\end{equation}
 
\noindent where $V_{\mathrm{WFB}}$ are the volumes of the uncertainty ellipsoids of the 
WFB sources in the PCs space calculated as 
$V_{\mathrm{WFB}}\!=\!\frac{4}{3}\pi\sigma_{\mathrm{PC}_{1}}\sigma_{\mathrm{PC}_{1}}\sigma_{\mathrm{PC}_{3}}$, 
and $V$ is the volume of the uncertainty ellipsoid in the 
PCs space of the generic WISE source considered. The logarithms of the 
volumes of the uncertainty ellipsoids are used to take into account the 
large number of order of magnitude potentially spanned by the differences 
between the volumes (always smaller than one in the PCs space though).
The above definition of the weighted score also has the effect of mapping 
the discrete distribution of scores calculated according to assignment law 
Equation~\ref{eq:score} into a continuous distribution that allows a finer 
classification of the candidate blazars.

\section{Selection of the candidate blazars}
\label{sec:candidates}

\begin{figure}[] 
      	\includegraphics[width=65mm]{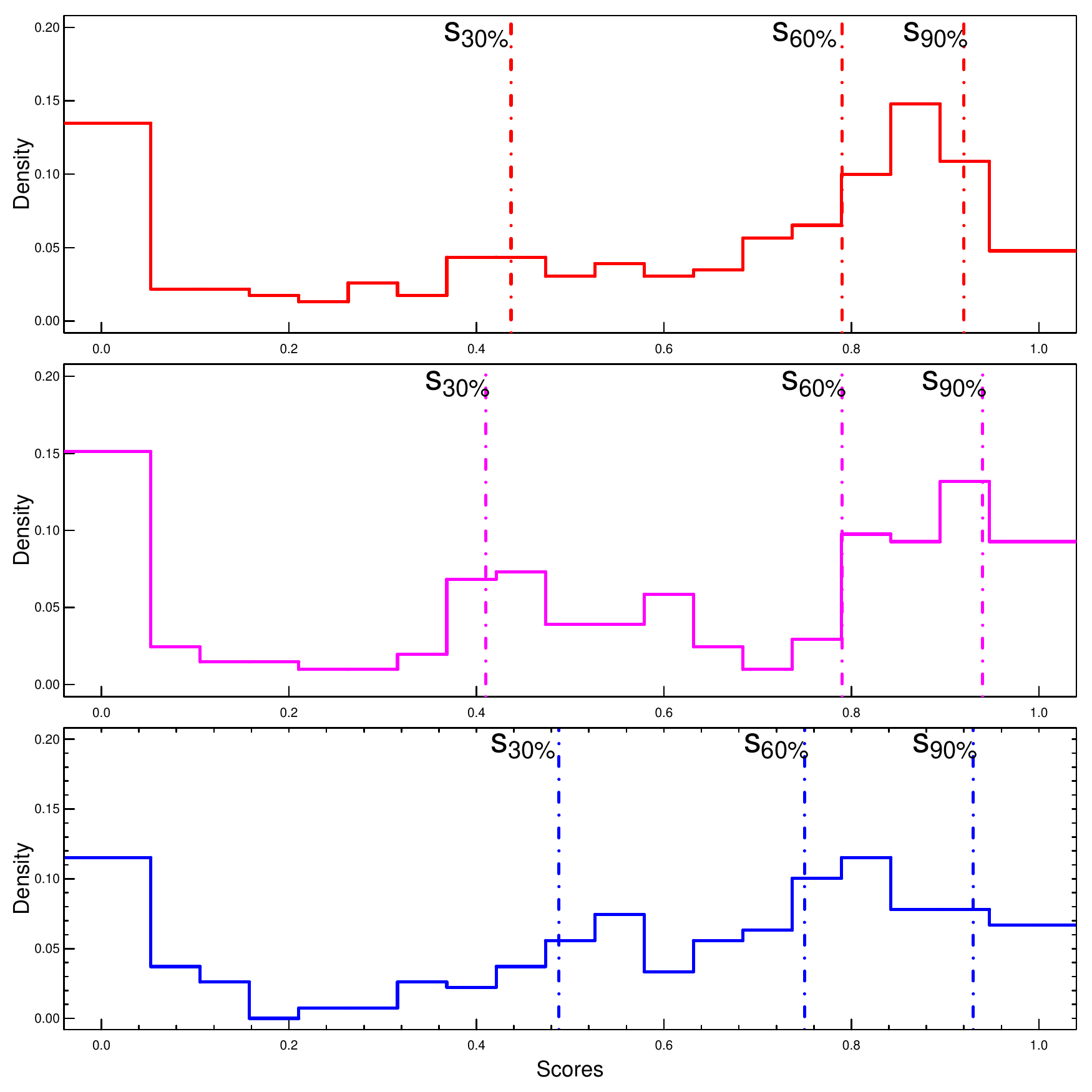}
          \caption{Histograms of distributions of score values calculated for the sources in the WFB sample 
          for the three regions of the \emph{locus} dominated respectively by the BZQs, the BZBs and
          in the mixed region (upper, mid and and lower panels respectively). The three 
          vertical lines in each panel represent $s_{30\%}$, $s_{60\%}$ and $s_{90\%}$. These 
          thresholds have been used to define the classes of candidate blazars (see text).}
          \label{fig:scoreswgs}
\end{figure}

The procedure for the evaluation of the scores based on the new parametrization of the 
WFB \emph{locus} discussed in the previous section is used to associate
high-energy sources to WISE candidate blazars. The WISE colors and their uncertainties for all 
the sources found in the WISE All-Sky photometry catalog within the region of positional 
uncertainty (hereinafter the Search Region - SR) of a given high-energy source and 
detected in all four WISE filters are retrieved, and the scores of these WISE sources are calculated
as described in Section~\ref{subsec:score}. Then, these sources are split among different classes 
according to the values of the their scores $s_{b}$, $s_{m}$ and $s_{q}$ for the BZB, Mixed 
and BZQ regions of the WFB \emph{locus} model
in the PCs space respectively. For each \emph{locus} region, every source 
is assigned to class A, class B, class C or is marked as an outlier based on its score values and relative 
to the threshold scores values defined as the 30\%, 60\% and 90\% percentiles of the distributions 
of scores in the three regions of the \emph{locus} for the WFB sources (see Figure~\ref{fig:scoreswgs}). 
The classes are sorted according to decreasing probability of the WISE source to be compatible with 
the model of the WFB \emph{locus}: class A sources are considered the most probable candidate 
blazars for the high-energy source in the SR, while class B and class C sources are less compatible
with the WFB \emph{locus} but are still deemed as candidate blazars. In more details, 
class A candidate blazars have score $s\leq s_{90\%}$, class B candidate blazars have score 
$s_{60\%}\!\leq\!s\leq\!s_{90\%}$ and class C candidate blazars have score 
$s_{30\%} \leq s\leq s_{60\%}$ for each region. The other sources considered 
outliers are discarded. The values of the score thresholds derived from 
the score distributions of WFB sources for the 
three regions of the \emph{locus} model are reported in Table~\ref{tab:thresholds} and shown 
in Figure~\ref{fig:scoreswgs} overplotted to the histograms of the score distributions of the WFB 
sources assigned to each of the three \emph{locus} regions.

\begin{table}[t]
	\begin{center}
	\caption{Values of the score thresholds $s_{30\%}$, $s_{60\%}$ and $s_{90\%}$, used for 
	the association experiments described in this proceedings. These values are determined as the 
	30\%-th, 60\%-th and 90\%-th percentiles of the scores of the WFB sample divided by BZB, 
	Mixed and BZB mixed regions.}
	\begin{tabular}{cccc}
	\\
	\tableline\tableline
					& BZB 	& Mixed	& BZQ  	\\
	\tableline
	$s_{30\%}$		& 0.48	& 0.44	& 0.41	\\
	$s_{60\%}$		& 0.75	& 0.79	& 0.79	\\
	$s_{90\%}$		& 0.93	& 0.92	& 0.94	\\	
	\tableline
	\label{tab:thresholds}
	\end{tabular}
	\end{center}
\end{table}

\noindent The choice of the percentiles used to define the classes of candidate blazars is 
arbitrary and can be changed to allow for more conservative (higher purity of the sample
of candidates) or more complete (lower purity of the sample of candidates) selections of 
candidate blazars in the SRs associated with unidentified high-energy sources. 

\subsection{Associations}
\label{sec:back}

\begin{center}
\begin{figure}[] 
	\includegraphics[width=65mm]{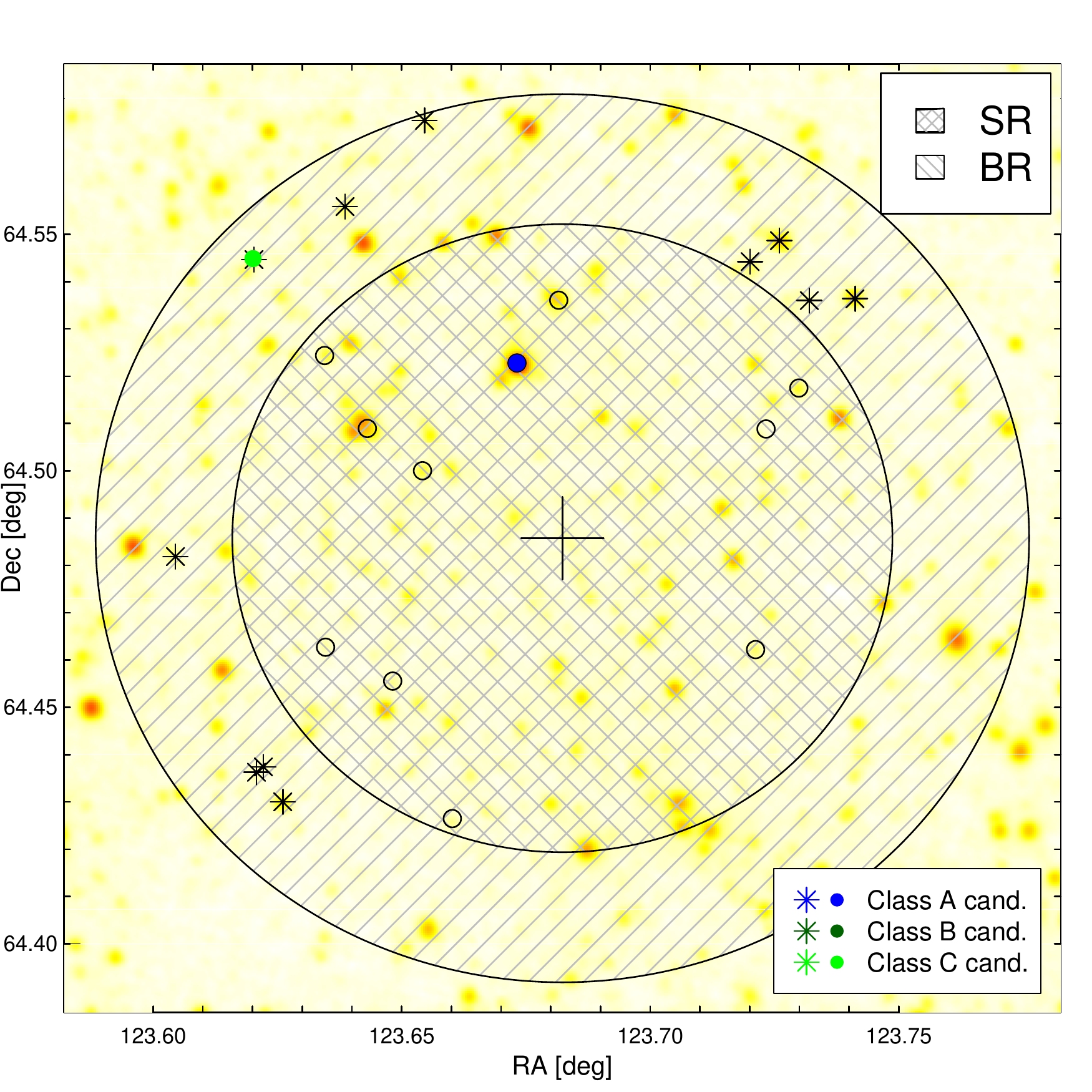}
           \caption{Results of the association procedure for a generic unassociated high-energy source
           superimposed on the image of the WISE sky around the position of the unassociated $\gamma$-ray
           source as seen in the $[3.4]\mu$m band. 
           The inner circle represents the Search Region (SR) of the high-energy source while the outer 
           circle delimits the annulus used as Background Region (BR). The open circles in the SR represent 
           the sources of the WISE All-Sky catalog 
           detected in all four WISE filters for which the scores have been evaluated (the sources not marked 
           by symbols in the image are not detected in at least one of the four WISE filters and have not been
           considered for the score evaluation). The solid circle
           represents the candidate blazar found within the SR and its color indicates that it is class A candidate
           blazar.}
          \label{fig:association}
\end{figure}
\end{center}

In our association procedure, the presence of WISE background sources with score values
that would qualify them as candidate blazars but that are not located within the SR of the  
unidentified high-energy source is taken into account by assessing
the number and type of spurious associations from sources within a local background 
region for each unassociated source.
For a generic SR of radius $r_{\mathrm{SR}}$, we define the background region (BR) as an 
annulus of outer radius $r_{\mathrm{BR}}\!=\!\sqrt{2}\!\cdot\!r_{\mathrm{SR}}$ and inner
radius equal to the SR radius and centered on the center of the SR. The SR and BR have same
area by definition. Within a given SR, all WISE sources detected in all four WISE filters are 
assigned a score value for each 
region of the \emph{locus} model, and successively ranked in classes using the same thresholds
used to classify the sources within the SR. An example of a generic SR and associated 
background region is shown in Figure~\ref{fig:association}, where the candidate blazar and 
the spurious BR candidate blazar are colored according to their class membership as defined in 
Section~\ref{sec:candidates}.

For every unassociated high-energy source, our method produces all candidate blazars 
(sources classified as class A, class B or class C candidate) in the SR. All candidate blazars 
located in the BR of the high-energy sources are also provided and can be used to evaluate 
the chance of spurious associations as a function of the class of the candidate blazars.

\section{Conclusions}

In this proceeding we have described the WFB sample of $\gamma$-ray emitting WISE blazars, 
gathered using the new \wse\ All-Sky release, the 2FGL catalog and the latest release of the \bzcat\
catalog. Then, we have presented a new association procedure for the unidentified high-energy sources
based on a new model of the \emph{locus} occupied by WFB sample in the three-dimensional
PCs space generated by the distribution of WFB \wse\ sources in the \wse\ color space. We defined
a quantitative measure of the compatibility of a generic \wse\ source with the \emph{locus} model
and expounded the new association procedure. This method can 
select candidate blazars classified as BZB or BZQ candidates and ranked according to the likelihood
of each candidate of being an actual blazar. We also investigated the possibility of spurious 
associations by determining the number and class of \wse\ sources compatible with the model of the 
WFB \emph{locus} in background regions defined around the SR of each high-energy source.

The performances of the method in terms of the efficiency and completeness have been estimated
in~\cite{dabrusco2013}, yielding a total efficiency $e_{\mathrm{tot}}\!\simeq\!97\%$ and total 
completeness $c_{\mathrm{tot}}\!\simeq\!81\%$ respectively. By using a $K$-fold cross-validation 
approach, we have also estimated the efficiency and 
completeness as functions of the \wse\ colors and galactic coordinates of the candidate blazars.

In~\citep{dabrusco2013}, we have presented the catalog of candidate blazars associated with 
the new procedure to the 2FGL $\gamma$-ray sources included in the WFB sample,
used to define the new model of the \emph{locus}. We have also discussed the catalog of 
candidate blazars obtained by applying the new association procedure to the 2FB sample, 
composed of all clean $\gamma$-ray sources associated with blazars in the 2FGL catalog 
but not contained in the WFB sample. We will make the code for the association and 
both catalogs of candidate blazars publicly available.

\bigskip 
\begin{acknowledgments}

The work is supported by the NASA grants NNX10AD50G, NNH09ZDA001N and 
NNX10AD68G. R. D'Abrusco gratefully acknowledges the financial 
support of the US Virtual Astronomical Observatory, which is sponsored by the
National Science Foundation and the National Aeronautics and Space Administration.
F. Massaro is grateful to A. Cavaliere, S. Digel, D. Harris, D. Thompson, A. Wehrle for 
their helpful discussions.
The work by G. Tosti is supported by the ASI/INAF contract I/005/12/0.
H. A. Smith acknowledges partial support from NASA/JPL grant RSA 1369566.
TOPCAT\footnote{\underline{http://www.star.bris.ac.uk/$\sim$mbt/topcat/}} 
\citep{taylor2005} and SAOImage DS9 were used extensively in this work 
for the preparation and manipulation of the tabular data and the images.
This research has made use of data obtained from the High Energy 
Astrophysics Science Archive
Research Center (HEASARC) provided by NASA's Goddard
Space Flight Center; the SIMBAD database operated at CDS,
Strasbourg, France; the NASA/IPAC Extragalactic Database
(NED) operated by the Jet Propulsion Laboratory, California
Institute of Technology, under contract with the National Aeronautics and Space Administration.
Part of this work is based on archival data, software or on-line services provided by the ASI Science Data Center.
This publication makes use of data products from the Wide-field Infrared Survey Explorer, 
which is a joint project of the University of California, Los Angeles, and 
the Jet Propulsion Laboratory/California Institute of Technology, 
funded by the National Aeronautics and Space Administration.

\end{acknowledgments}

\bigskip 

\end{document}